    \documentclass[prl,aps,twocolumn,showpacs]{revtex4}
\usepackage[dvips]{graphicx}
\usepackage{dcolumn}%
\usepackage{amsmath}%
\usepackage{amsfonts}%
\usepackage{amssymb}
\providecommand{\U}[1]{\protect\rule{.1in}{.1in}}
\newcommand{\be}{\begin{equation}}
\newcommand{\ee}{\end{equation}}
\newcommand{\bea}{\begin{eqnarray}}
\newcommand{\eea}{\end{eqnarray}}
\newcommand{\bt} {\begin{tabular}}
\newcommand{\et} {\end{tabular}}
\newcommand{\nn}{ \nonumber}
\newcommand{\ds}{\displaystyle}
\newcommand{\ba} {\begin{align}}
\newcommand{\ea} {\end{align}}
\begin{document}

\title{Electromagnetic quantum waves and their effect on the low temperature magnetoacoustic response of a quasi-two-dimensional metal}
\author{Natalya A. Zimbovskaya}
\affiliation{Department of Physics and Electronics, University of Puerto , 
Rico-Humacao, CUH Station, Humacao, Puerto Rico 00791,  USA}
\affiliation{Institute for Functional Nanomaterials, University of Puerto Rico, San Juan, Puerto Rico 00931, USA}

\begin{abstract}
 We theoretically analyze  weakly attenuated electromagnetic waves in quasi-two-dimensional (Q2D) metals in high magnetic fields. Within the chosen geometry, the magnetic field is directed perpendicularly to the conducting layers of a Q2D conductor. We showed that longitudinal collective modes could propagate along the magnetic field provided that the Fermi surface is moderately corrugated. The considered waves speeds strongly depend on the magnetic field magnitude. Also, we analyzed interactions of these quantum waves with sound waves of fitting polarization and propagation direction, and we showed that such interaction may bring significant changes to the low temperature magnetoacoustic response of Q2D conductors.   
   \end{abstract}

\date{\today}
\maketitle

\subsection{\normalsize I. Introduction}

As known for a long time, electromagnetic waves  incident at a surface of a metal from outside, cannot penetrate into the bulk of  material deeper than a thin layer adjoining the surface (skin layer) \cite{1}.  
 The suppression of the electric field inside the metal is caused by the response of conduction electrons. 
It occurs when the incident wave frequency $ \omega $ is smaller than the charge carriers plasma frequency $ \omega_0 ,$ which determines the characteristic time of their response to the electromagnetic disturbance. When $ \omega > \omega_0, $ the conduction electrons do not have sufficient time to respond, so the  metal becomes transparent for electromagnetic waves of high frequencies. Here, we concentrate on the waves whose frequencies belong to the range $ \tau^{-1} < \omega \ll \omega_0 $ where $ \tau^{-1} $ is the characteristic frequency of the conduction electrons scattering by various types of scatterers  (such as phonons and/or lattice imperfections).  When the wave frequency  remains within the considered range, some charge carriers move along the direction of the wave propagation with the speed coinciding with the phase velocity of the wave. At weak scattering $(\omega\tau > 1)$ such quasiparticles may stay ``in phase" with the wave during several wave periods. All this time they see the electric field of the same magnitude and direction, and the field accelerates them. As a result, the electric field energy is transferred to the conduction electrons thus weakening the field.

An external magnetic field $ \bf B $ applied to the metal affects electron motions in the planes perpendicular to its direction. The effect becomes significant in high magnetic fields when the cyclotron frequency of electrons $ \Omega $ exceeds their scattering frequency $ 1/\tau. $ Under such conditions, so called windows of transparency appear in the $ {\bf q},\omega $ space $ ({\bf q},\omega $ are the wave vector and the frequency of the electromagnetic wave, respectively). These are regions  where the above described collisionless absorption of the external electromagnetic field cannot occur. As a result, various kinds of weakly attenuated  electromagnetic waves 
whose frequencies and wave vectors belong to the regions of transparency  (such as helicoidal, cyclotron and magnetohydrodynamic waves) may occur in metals \cite{2,3}. 
All these waves could be excited by an external electromagnetic field, and they are generated by collective motions of conduction electrons. At low temperatures $ T \ (\hbar\Omega \gg kT,\ k $ being the Boltzmann's constant)  the quantization of the electron motions in the planes perpendicular to the field $ \bf B $ appears. 
 This gives rise to oscillations of the electron density of states at the Fermi surface (FS) of a metal. These quantum oscillations generate several effects such as de Haas - van  Alphen oscillations in the magnetization and Shubnikov - de Haas oscillations in the  magnetoresistivity. Also, the quantization of the conduction electrons motions brings changes into the geometry of transparency regions, creating opportunities for occurrence of particular modes which 
 cannot exist  otherwise.

To explain this we assume that a quantizing magnetic field is applied along the ``z" axis of the chosen coordinate system $\big ({\bf B} = (0,0,B)\big) $. Then the conduction electrons acquire Landau energy spectra of the form: 
  \be
 E(n,p_z,\sigma) = \hbar\Omega(n + 1/2) + E_{||} (p_z,\sigma)   \label{1}
  \ee
   where $p_z $ is the electron momentum projection on the magnetic field direction, the quantum number $ n $ labels Landau levels, $ \sigma $ is the spin quantum number, and the term $ E_{||} (p_z,\sigma) $ originates from the electron motion along the magnetic field. As follows from the Eq. (\ref{1}), at a certain value of the magnetic field the ``longitudinal" part of the energy  of an electron at the FS $ E_{||}(p_z,\sigma) $ can take on values, belonging to a set of rather narrow intervals whose widths are of the order of thermal energy $ kT. $ This leads to a similar distribution of possible values of the longitudinal velocity $ v_z = \partial E_{||}/\partial p_z $ at the Fermi surface. Within the low temperature limit $ (T\to 0), $ these intervals are reduced to the points, and we obtain a discrete set of possible $ v_z $ values for every given $ B. $ Let us assume that the wave travels along the magnetic field $ \big({\bf q} = (0,0,q)\big) ,$ and the condition $ qv_z/\Omega \ll 1 $ is satisfied. Under this condition, the charge carriers displacements in ``z" direction during the cyclotron period $ 2\pi/\Omega $ are much smaller than the wavelength $ \lambda = 2\pi/q. $ Therefore, quasiparticles are seeing nearly the same electric field after each turn over their cyclotron orbits. They may efficiently absorb the wave energy provided that at least one among the permitted values of the longitudinal velocity equals the phase velocity of the wave. Due to the fact that quantization of the charge carriers motions puts essential restrictions on the possible values of $ v_z, $ extra regions of transparency appear thus creating opportunities for the specific modes to occur. Possible appearance of such  weakly damped modes in quasi-isotropic metals was predicted by Konstantinov and Perel \cite{4}, and they were named ``quantum waves" \cite{5}. Subsequently, their spectra and coupling with ultrasonic waves traveling in a metal were theoretically analyzed in several works \cite{6,7,8}. However, this analysis showed that experimental observation of quantum waves in conventional three-dimensional metals should be very difficult, because extremely low temperatures are required for their appearance.

In the last three decades various layered conducting materials  were synthesized. These are organic conductors belonging to the family of tetrathiafulvalene salts, dichalcogenides of transition metals, graphite and its intercalates, cuprates and some other. Strong anisotropy of the electrical conductivity is inherent for all these materials. The anisotropy originates from the fact that conducting layers are weakly coupled to each other. Correspondingly, a charge carrier energy is nearly independent of the momentum  perpendicular to the layers, and the charge carriers spectra are quasi-two-dimensional (Q2D). Fermi surfaces of Q2D conductors could be described as systems of weakly corrugated cylinders \cite{9,10,11}. These materials attract significant interest of the research community, and their electron characteristics are being intensively studied both theoretically and experimentally. In particular, it was shown that the specific  FS geometry inherent for Q2D conductors provides better opportunities for occurrence and observation of weakly attenuated electromagnetic waves, which could appear in strong but nonquantizing magnetic fields \cite{12}.      
 In the present work we theoretically analyze spectra of longitudinal ``quantum waves'' in Q2D metals and their interaction with sound waves. 

\subsection{\normalsize II. Dispersions of longitudinal quantum waves}

In our analysis we assume that the field $ {\bf B} $ is orthogonal to the conducting layers, and we use a simple tight-binding approximation for the charge carriers dispersion:
   \be
E({\bf p}) = \frac{{\bf p}_\perp^2}{2m_\perp} - 2t\cos \left(\frac{\pi p_z}{p_0}\right) \equiv \frac{{\bf p}_\perp^2}{2m_\perp} - 2 t\cos (k_z d)  \label{2}
  \ee
 Here, $ \bf p_\perp $ is the momentum projection on the layer plane and $ m_\perp $ is the effective mass corresponding to the motion in this plane, respectively. The parameter $ t $ in the equation (\ref{2}) is the interlayer transfer integral, and $p_0 = \pi\hbar / d $ where $ d$ is the distance between the adjacent layers. In a quantizing magnetic field, the charge carriers spectra are described by the Eq. (\ref{1}) with the longitudinal energy of the form:
  \be
E_{||}(p_z,\sigma) = - 2t\cos\left(\frac{\pi p_z}{p_0} \right) + \frac{\sigma g}{2} \beta_0 B              \label{3}                    \ee
   where $ g $ is the spin-splitting coefficient and $ \beta_0 $ is the Bohr's magneton. 
  
In general, spectra of weakly attenuated waves in metals are determined by the dispersion equation of the form \cite{1,2}:
  \be
 \mbox{Det} \left[q^2 \delta_{ij} - q_i q_j - \frac{\omega^2}{c^2} \epsilon_{ij} (\omega, {\bf q}) \right] = 0.  \label{4}
  \ee
  Here, $q_i,q_j $ are the components of the wave vector $ \bf q $ within the chosen coordinate system, $ \delta_{ij} $ is the Kronecker symbol and the 
  dielectric permittivity tensor $ \epsilon_{ij}(\omega,\bf q)$ is related to the conductivity tensor $ \sigma_{ij}(\omega,\bf q):$
  \be
 \epsilon_{ij} (\omega,{\bf q}) = \delta_{ij} + \frac{4\pi i}{\omega c^2} \sigma_{ij} (\omega,\bf q)                   \label{5}                         \ee
  So, the last term in the dispersion equation describes the response of a metal to an electromagnetic disturbance.          The electric conductivity tensor is  defined by the relation:
  \be
 J_{q\omega}^\alpha = \sigma_{\alpha\beta} (\omega, {\bf q}) E_{q\omega}^\beta    \label{6}
  \ee
 where ${\bf J}_{q\omega} $ and  ${\bf E}_{q\omega} $ are the Fourier transforms of the charge carriers current density and the electric field, respectively. To compute the current density, one needs to find the expressions for matrix elements of the charge carriers density matrix. When the external disturbance of the conduction electrons system is weak, the density matrix includes the equilibrium term and small nonequilibrium correction $ \rho ,$ describing the linear response of the system to the disturbance. The density matrix $ \rho $ satisfies the equation:
  \be
 i\hbar \frac{\partial \rho}{\partial t} = [H,\rho ]   \label{7}
  \ee 
  where $ H = H_0 + H_1 $ is the single-particle Hamiltonian for charge carriers exposed to  an external disturbance. In the considered case, the Hamiltonian $ H_0$  eigenvalues $E_\nu$ are the energies given by the Eqs. (\ref{1}), (\ref{3}), and the term $ H_1 $ describes the effects of the electromagnetic disturbance and scattering. The kinetic equation (\ref{6}) may be rewritten in the form \cite{13,14}:
  \be
 \frac{1}{i\hbar} (E_{\nu'} - E_{\nu} + \hbar\omega)\rho_{\nu\nu'} + \frac{1}{i\hbar} (f_\nu - f_{\nu'}) W_{\nu\nu'} = I_{\nu\nu'} [\rho]   \label{8}
  \ee
  where $f_\nu $ is the Fermi distribution function for the energy $ E_\nu $ and $ I_{\nu\nu'}[\rho]$ is the collision integral which describes the charge carriers scattering.  Considering a high frequency disturbance $(\omega\tau > 1) $ we may use the simple approximation:
  \be
 I_{\nu\nu'} [\rho] = - \frac{1}{\tau} \rho_{\nu\nu'}  . \label{9}
  \ee 
 For simplicity, we omit from consideration the effects arising due to interactions between the charge carriers. Then the term $W_{\nu\nu'} $ solely corresponds to the effect of the electromagnetic disturbance and can be presented as follows:
  \be
 W_{\nu\nu'} = en_{\nu\nu'} (-{\bf q}) \Phi_{\bf q \omega} + \frac{i\omega}{c} {\bf j}_{\nu\nu'} ({-\bf q}) A_{\bf q\omega}   \label{10}
 \ee
 where $en_{\nu\nu'} ({-\bf q})  $ and $ {\bf j}_{\nu\nu'} ({-\bf q}) $ are the Fourier transforms of matrix elements of the operators of charge and current densities, respectively, and $ \Phi_{\bf q \omega} $ and $ \bf A_{q\omega} $ are the Fourier transforms of the scalar and vector potentials of the electromagnetic field  disturbing the conduction electrons.  The current density ${\bf J}_{q\omega} $ equals:
  \be
 {\bf J}_{q\omega} =\sum_{\nu\nu'} \rho_{\nu\nu'}  (\omega,\bf q) j_{\nu'\nu} (q).  \label{11}
  \ee
 Solving the kinetic equation (\ref{8}) and substituting the results for $ \rho_{\nu\nu'} (\omega, \bf q) $ into Eq. (\ref{11}) we may find  expressions for the conductivity components.

   Below, we consider longitudinal modes traveling along the magnetic field 
 so the dispersion equation (\ref{4}) is reduced to the form:
  \be
 \epsilon_{zz} (\omega,{\bf q}) = 0.   \label{12}
 \ee
  Employing Eqs. (\ref{8})-(\ref{11}) and the continuity equation for the electric charge:
  \be
 \frac{e}{i\hbar} (E_{\nu'} - E_\nu) n_{\nu\nu'} ({\bf -q})= i \bf q j_{\nu\nu'} (-q)  \label{13}
  \ee
  we may get the following expression for relevant conductivity component:
      \be  
  \sigma_{zz}(\omega,{\bf q}) =  - \frac{ie^2\omega}{q^2} \Big(1 + \frac{i}{\omega\tau}\Big)
   \sum_{\nu\nu'} \frac{(f_\nu - f_{\nu'}) n_{\nu\nu'} ({\bf -q}) n_{\nu'\nu}\bf (q)}{E_{\nu'} - E_\nu + \hbar\omega + i\hbar/\tau}  . \label{14}
  \ee 
As shown in the Appendix the expression for $\sigma_{zz} $ may be converted to the form:
    \begin{align}
 & \sigma_{zz}(\omega,{\bf q}) =  - \frac{i\omega e^2}{q^2} (Y_+ + Y_-)
  \nn\\ = & \tilde\sigma ({\bf q}) u \big\{\pi u + \Gamma_+ + \Gamma_- - i(1 + \eta_+ + \eta_- + \Delta_+ + \Delta_-)  \big\}  \label{15}
   \end{align}
  where
 \be
 \tilde\sigma ({\bf q}) = \frac{\pi Ne^2m_\perp v_0}{q A} ,
 \qquad  u = \frac{\omega}{qv_0} ,  \label{16}
  \ee
 $ A $ is the mean cross-sectional area of the FS,  $ v_0  = 2\pi t/p_0 $ is the maximum  value of  the charge carriers velocity component along the magnetic field,  $ \pm $ correspond to the different spin orientations with respect to the magnetic field  
 and $ N $ is the charge carriers density. The Eq. (\ref{15}) includes both real and imaginary parts.
 It gives a good approximation for the considered conductivity component at small values of $ u\ (u \ll 1). $
  The real part presents the conductivity as such whereas the imaginary part describes the phase shift of the electric field of the incident wave. 

   The oscillating terms $ \eta_\sigma $ and $ \Gamma_\sigma $ are described by the real and imaginary parts of the function $ X_\sigma $ which has the form:
   \begin{align}
X_\sigma =  \frac{1}{2qR} \sum_{r=1}^\infty \frac{(-1)^r}{r} D(r) \exp\left[-2\pi i r \left(\frac{F}{B} - \frac{\sigma g}{2} \frac{\Omega_0}{\Omega}\right) \right]
  \nn \\   \times  
\left \{\cos \left[\frac{4\pi rt}{\hbar\Omega} \sqrt{1 - U_+^2} \right]  
- \frac{|U_-|}{U_-} \cos \left[\frac{4\pi rt}{\hbar\Omega} \sqrt{1 - U_-^2}\right] \right \}.
                 \label{17}
                                   \end{align} 
    Here $ U_\pm = (u \pm \pi\hbar q/2p_0), \  \hbar \Omega_0 = \hbar \beta_0 B $ is the spin splitting energy, $ R = qv_0 /\Omega,\ F = cA/2\pi\hbar e .$  The damping factor $ D(r) = R_T(r) R_D (r) $ describes deterioration of the magnetic oscillations due to the effect of temperature $ (R_T (r)) $ and charge carriers scattering $ (R_D(r)).$

Additional oscillating corrections $\Delta_\sigma, $ which appear in the Eq. (\ref{15}),  are closely related to the magnetic quantum oscillations of the charge carriers density of states on the Fermi surface. When the latter is noticeably warped $ (t/\hbar\Omega > 1) $ and the cyclotron quantum $ \hbar\Omega $ is much smaller than the charge carriers chemical potential $ \mu \ (u\sqrt{\mu/\hbar\Omega} > 1) ,$ the terms $ \Delta_\sigma $ may be approximated as follows \cite{15}:
    \begin{align}
   \Delta_\sigma = & \sqrt{\frac{\hbar\Omega}{2\pi^2t}} \sum_{r=1}^\infty 
\frac{(-1)^r}{\sqrt r} D(r) 
      \nn \\  & \times
  \left\{\cos\left[\frac{2\pi r F_{max}}{B} - \pi r \sigma \frac{ g}{2} \frac{\Omega_0}{\Omega} - \frac{\pi}{4} \right] \right.
  \nn \\ & +  \left.
\cos\left[\frac{2\pi r F_{min}}{B} + \pi r \sigma \frac{g}{2} \frac{\Omega_0}{\Omega} + \frac{\pi}{4} \right] \right\}.   \label{18}
   \end{align}
 In this expression, $ R_D(r) = \exp(-\pi r/\Omega\tau), $ so the effect of scattering may be accounted for by replacing the temperature $ T $ in the expression for $ R_T (r) = ry/\sinh (ry) \ (y = 2\pi^2 kT/\hbar\Omega) $ by an ``effective temperature" $T^* = T + T_D $ where $ T_D = \hbar/2\pi k\tau $ is the Dingle temperature \cite{16}.

Now, we compare two terms in the expression (\ref{5}),  and we conclude that in the case of $ \epsilon_{zz},$ the second term significantly exceeds the first one provided that $ \omega \ll \omega_0. $ We omit the lesser term and  reduce the dispersion equation (\ref{12}) to a very simple form:
    \be 
  Y_+ + Y_- = 0   \label{19}
  \ee         
  When the FS becomes perfectly cylindrical $(t \to 0),$ the difference between the terms included  in the Eq. (\ref{17}) disappears, and the oscillating functions $ \eta_\sigma $ and $ \Gamma_\sigma $ turn zero. This agrees with the results of a recent work \cite{15}, where the analysis was carried out in the limit of small $ q. $ So, the dispersion equation (\ref{19}) does not have solutions corresponding to weakly attenuated waves when 2D conductors are considered. This conclusion seems  reasonable because electromagnetic waves in metals are generated by collective motions of conduction electrons. At $ t = 0$ the charge carriers cannot move between the conducting layers of a layered conductor, and this prevents the occurrence of a longitudinal collective mode propagating along the perpendicular to the layers.
   Such modes could appear in Q2D metals with moderately corrugated Fermi surfaces, where the inequality $ t > \hbar\Omega $ may be satisfied in strong magnetic fields.  To prove this statement, we analyze the behavior of the  oscillating terms $ \eta_\sigma $ and $ \Gamma_\sigma. $

  As discussed in the introduction, the quantization of the charge carriers motions puts restrictions on the allowed values of the longitudinal component of the their velocity $ v_z = \partial E/\partial p_z $ at the Fermi surface. In the low temperature and collisionless limit, the $ v_z $ values at a certain magnitude of the field $ \bf B $ belong to a discrete set. For a Q2D FS determined by the Eq. (\ref{2}), this set consists of two subsets related to the FS cross-sections with the maximum $\{u_{n,\sigma}\} $ and minimum $\{u_{m,\sigma}\} $ cross-sectional areas. 
  Using the Eqs. (\ref{1}), (\ref{3}) which determine the charge carriers spectra in Q2D conductors in the presence of a quantizing magnetic field, we may derive expressions for the permitted longitudinal velocities of the quasiparticles at the Fermi surface:
   \be
 \frac{\hbar\Omega}{t} \Big(\frac{F_{\max}}{B} - \frac{\sigma g \Omega_0}{\Omega} - n - \frac{1}{2} \Big) = 2\Big(1 - \sqrt{1 - w_{n,\sigma}^2}\, \Big) ,          \label{20}
  \ee 
 \be
\frac{\hbar\Omega}{t} \Big(m  + \frac{1}{2} - \frac{F_{\min}}{B} - \frac{\sigma g \Omega_0}{\Omega} \Big) = 2\left(1 - \sqrt{1 - v_{m,\sigma}^2}\, \right)             \label{21}
   \ee
  where $ w_{n,\sigma} = u_{n,\sigma}/v_0 $ and $ v_{m,\sigma} = u_{m,\sigma}/v_0. $
  It is shown in the Appendix that the functions $ \eta_\sigma $ diverge when $ U_\pm $ approaches one of the  dimensionless velocities introduced by Eqs. (\ref{20}), (\ref{21}). 
 This result indicates that the dispersion equation (\ref{19}) may have solutions corresponding to the considered quantum waves.

Under practical conditions, the combined influence of temperature and charge carriers scattering leads to the smoothing over the divergencies. To estimate their effect on the $\eta_\sigma $  oscillations, we again use the well known expression $ R_T(r) = ry/\sinh(ry) $ where $ y = 2\pi^2kT/\hbar\Omega.  $ Also, we approximate the scattering factor as $ R_D(r) = \exp (-\pi r u t /\hbar\Omega ql) $. Then we obtain 
   \be
 |\eta_\sigma| \sim u \frac{\hbar\Omega}{kT} \zeta \left (2; \frac{b+1}{2}\right)   \label{22}
     \ee
  where $ \ds b = \frac{t}{kT} \frac{u^2}{(\omega\tau)^2}, $ and $ \zeta $ is
Riemann's zeta function. As follows from Eq. (\ref{22}),  at $ \omega\tau > u \sqrt{t/kT} > 1 $ the magnitude of the oscillations is determined by the temperature $(|\eta_\sigma| \sim u\hbar\Omega/kT)$ whereas at  $ 1< \omega \tau< u\sqrt{t/kT} $ the effect of charge carriers scattering predominates $(|\eta_\sigma| \sim ql\hbar\Omega/t).$ Now, we combine these results keeping in mind that oscillating terms given by Eq.(\ref{17}) must be taken into consideration  when the Fermi surface of a layered conductor is moderately corrugated $(t > \omega p_0/q ).$ This follows from the results presented in the Appendix. So, we conclude that the function $ \eta_\sigma $ may reach values exceeding unity at the peaks of its quantum oscillations provided that:
   \be
  u \frac{\hbar\Omega}{kT} > 1; \qquad ql\frac{\hbar\Omega}{t} > 1.   \label{23}
  \ee
  where $ l = v_0 \tau. $ When these conditions are satisfied, the dispersion equation (\ref{19}) has solutions corresponding  to the quantum waves.
   As follows from the dispersion equation, the imaginary terms in the Eq. (\ref{21}) determine the quantum waves attenuation. 

The behavior of the functions $ \eta_\sigma $ and $ \Gamma_\sigma $ in the low temperature and weak scattering region defined by the inequalities (\ref{21})  is shown in the Fig. 1. In plotting the curves presented in this figure we used the relevant data reported for a typical Q2D organic metal$\beta-(ET)_2IBr_2 $ \cite{11}, namely: $ A = 1.31 \times 10^{-49}(kgms^{-1})^2,\ m_\perp = 4.5 m_0\  (m_0 $ being a free electron mass) and $ (A_{\max} - A_{\min})/A \sim 0.04. $ Using these results we estimate the velocity $ v_0 $ and the transfer integral $ t, $ and we get $ v_0 \sim 4 \times 10^3 ms^{-1} $ and $ t \sim 2meV, $ respectively. As shown in the Fig. 1, contributions to $ \eta_\sigma $ from both maximum and minimum cross-sectional areas of the Q2D Fermi surface reach values exceeding unity at the peaks of oscillations at $ T \sim 10 mK. $ Presently, such temperatures are quite accessible in experiments. As for the  term $ \Gamma_\sigma, $ its magnetic field dependence reveals a sequence of nearly rectangular  peaks slightly smoothed near the vertexes due to the effects of temperature and scattering. The peaks heights and widths are determined by the value of the parameter $ qR. $ In the limit of long wavelengths $ (qR \ll 1),$ the peaks become very high and narrow similar to those describing so called giant quantum oscillations in the ultrasound attenuation observed in some conventional 3D metals \cite{1}.  
   In the present case, these peaks appear as a result of collisionless absorption of the electromagnetic waves energy by charge carriers. The effect of giant quantum oscillations of electromagnetic field attenuation was predicted for quasi-isotropic metals in the works \cite{17,18}. 
The peaks are separated by the intervals where $ \Gamma_\sigma $ equals $-\pi u/2 $ and appears completely balanced by the semi-classical contribution to the imaginary part of $ Y_\sigma. $

 \begin{figure}[t]  
\begin{center}
\includegraphics[width=4.6cm,height=9cm,angle=-90]{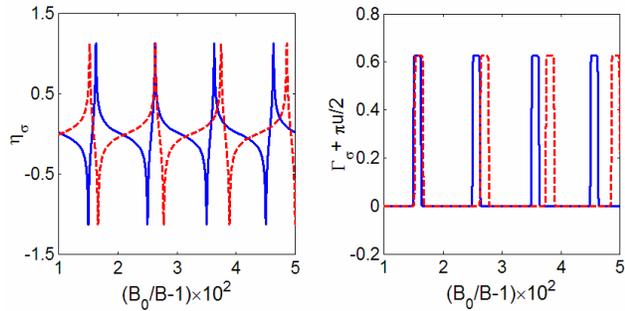}
\end{center}
\caption{
(Color online) Magnetic quantum oscillations in the real (left panel) and imaginary (right panel) parts of the function $ Y_\sigma. $ Solid and dashed lines correspond to the contributions from minimum and maximum cross-sections of a moderately corrugated Fermi surface of a Q2D conductor. The curves are plotted assuming that $ F_\sigma/B_0 = 100,\ t/\hbar\Omega(B_0) = 4,\ T = 10mK,\ B_0 = 20 T,\ m_\perp = 4.5 m_0\ $.
}%
 \label{rateI}%
\end{figure}

To further clarify the physical meaning of $ \Gamma_\sigma $ we turn to the expression (\ref{14}) for $ \sigma_{zz} (\omega,\bf q). $
  We remark that the condition of vanishing of the imaginary part agrees with the dispersion equation 
for the undamped quantum waves. The dependencies of the real and imaginary parts of $ \sigma_{zz} $ on the frequency $ \omega $ at a certain fixed value of the magnetic field are shown in the Fig. 2. The real part exhibits peaks within some frequency intervals and remains zero beyond them. Within these regions of high conductivity charge carriers intensively absorb the energy of the incident wave preventing its propagation in the metal.
 The peaks in the conductivity appear as a result of summing up contributions from $ \Gamma_{+} $ and $ \Gamma_{-} .$ At low temperatures both $\Gamma_\sigma $ reveal sequences of nearly rectangular peaks. The relative positions of nearest peaks in $ \Gamma_{+} $ and $\Gamma_{-} $ are determined by the magnitude of the magnetic field $ B ,$ the value of the transfer integral $ t ,$ which controls the difference between the maximum and minimum cross-sectional areas of the FS, frequency $ \omega $ and spin splitting characterized by  $ g \frac{\Omega_0}{\Omega}. $ Varying these parameters one may observe subsequent variations in the relative positions of the nearest peaks. At some values of the relevant parameters th nearest peak $ \Gamma_{+} $ and $ \Gamma_{-} $ are completely or partially overlapped, at order values they stay apart from each other. As a result, the conductivity dependence of frequency may reveal moderately low trapezoid-like peaks corresponding to separate nearest peaks in $ \Gamma_\sigma $ for different spin orientations or higher features of a characteristic bottle-like shape, which appear due to the partial overlapping of the nearest peaks in $ \Gamma_\sigma .$ Such features are shown in the Fig. 2. 
 
 High trapezoid-like peaks occurring, which the nearest peaks in $ \Gamma_{+} $ and $ \Gamma_{-} $ completely overlap may appear in the frequency dependence of Re$ \sigma_{zz}, $ as well.   
 Longitudinal quantum waves may propagate in the regions of low conductivity where the metal becomes ``transparent" for the electromagnetic fields.
The locations of the regions of transparency are determined by relations:
   \begin{align}
  & w_{n+1,\sigma}^2 < U_+^2 < w_{n,\sigma}^2; \ \qquad 
   w_{n'+1,\sigma'}^2 < U_-^2 < w_{n',\sigma'}^2;
  \nn \\ 
  &   v_{m,\sigma}^2 < U_+^2 < v_{m+1,\sigma}^2; \  \qquad 
v_{m',\sigma'}^2 < U_-^2 < v_{m' +1,\sigma'}^2. \label{24}
 \end{align}  
  These inequalities may be consistently satisfied when Landau  quantum numbers  $ n,n',m,m' $ and spin numbers $ \sigma,\sigma' $ accept fitting values. They specify the regions where the dispersion equation (\ref{19}) has solutions corresponding to weakly attenuated longitudinal modes propagating along the magnetic field $ \bf B. $

 \begin{figure}[t]  
\begin{center}
\includegraphics[width=4.6cm,height=9cm,angle=-90]{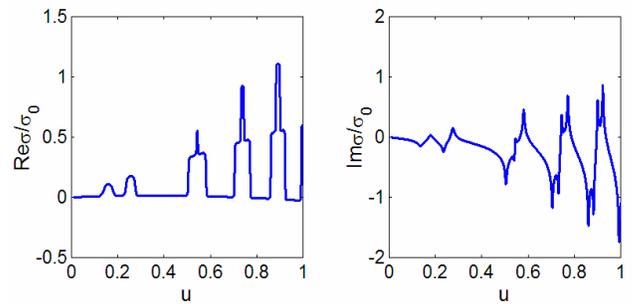}
\end{center}
\caption{
(Color online) The dependencies of the real (left panel) and imaginary (right panel) parts on $ \sigma_{zz} $ of $ \omega/qv_0. $ The curves are plotted neglecting the spin-splitting of the charge carriers energy levels to simplify the shape of the peaks in $ \sigma_{zz} $ real part. However, the contributions form both maximum and minimum cross-sectional areas are included. The curves are plotted assuming $ B = B_0 = 20 T  $ and using the same values for the remaining parameters as in the Fig. 1. 
}%
 \label{rateI}%
\end{figure} 

19
Now, we analyze the effect of the corrections $ \Delta_\sigma $ on the dispersions of the quantum waves. The magnitude of $ \Delta_\sigma $ at the peaks of oscillations may be estimated as:
   \be    
|\Delta_\sigma| \sim \left(\frac{\hbar\Omega}{2\pi^2t}\frac{\hbar\Omega}{kT^*} \right)^{1/2}.         \label{25}                                   \ee 
      In principle, $ \Delta_\sigma $ magnitude may reach values compared to unity when the effective temperature $ T^* $ becomes extremely low thus providing negligibility of thermal and scattering effects. However, this may happen at significantly lower temperatures and weaker scattering than those needed for the terms $ \eta_\sigma $ to reach values of the order of unity at the oscillations peaks. To justify this statement, we take the same values for the parameters characterizing the Fermi surface which were used in plotting of the Fig. 1, namely:  $ B = 20T,\  \omega \approx 10^{12} s^{-1} , $  $ \omega\tau \sim 1, $ and $ T= 10 mK. $ This gives for the Dingle temperature the estimation $ T_D \sim 100 mK, $ thus providing that peak values of $ \Delta_\sigma $ are approximately ten times smaller that the corresponding values of $ \eta_\sigma $ and significantly smaller than unity. On these grounds, we omit the corrections  $ \Delta_\sigma $ in further analysis.

20
Nevertheless, it is worthwhile to remark that the charge carriers scattering differently affects $ \eta_\sigma $ and $ \Delta_\sigma. $ This difference reflects the diversity in the nature and origin of these oscillating contributions to $ Y_\sigma. $ The terms $ \Delta_\sigma $ appear solely due to quantization of the charge carriers motion in the planes perpendicular to the strong magnetic field. The charge carriers motions along the magnetic field direction do not noticeably modify these  terms. Naturally, the effect of scattering is completely determined by the value of $ \Omega\tau, $ which indicates how long a charge carrier could stay at a certain cyclotron orbit. On the contrary, charge carriers motions along the magnetic field are crucial for occurrence of the terms $ \eta_\sigma. $  Therefore, the effect of scattering is mostly specified by the value of $ ql $ which indicates how long a charge carrier could preserve 
 its motion ``in phase" with the electromagnetic wave absorbing its energy.

21
We can easily  solve the dispersion equation (\ref{19}) within the long wavelength limit $ qR \ll 1. $ In this case, the inequalities  (\ref{24}) determining the regions of transparency for the considered waves, may be reduced to the form:
   \be
 w_{n+1,-} < u < w_{n,+}; \qquad v_{n,+} < u< v_{n+1,-},  \label{26}
  \ee
  where $ \pm $ indicate two spin orientations. The terms $ \eta_\sigma $ given by Eq. (\ref{17}) may be approximated as follows: 
  \begin{align}
  \eta_\sigma = & \frac{\pi}{4} u \left\{ \cot \left[2\pi \left( \frac{F_{\max}}{B}  - \frac{\sigma g}{2} \frac{\Omega_0}{\Omega} - \frac{t}{\hbar\Omega}u^2 \right)\right] \right.
  \nn \\ & +  \left.
 \cot\left[2\pi \left( \frac{F_{\min}}{B}  + \frac{\sigma g}{2} \frac{\Omega_0}{\Omega} + \frac{t}{\hbar\Omega}u^2 \right)\right] \right\}.    \label{27}
   \end{align}
 In the expression (\ref{27}), the cotangents diverge when $ u $ approaches one of the velocities $ w_{n,\sigma} $ and/or $ v_{n,\sigma}.$ As follows from Eqs. (\ref{19}), (\ref{27}), within the long wavelength limit, the weakly attenuated longitudinal modes determined by the Eq. (\ref{19}), have linear dispersions, and their speeds strongly depend of the magnetic field magnitude, being close to $ w_{n,\sigma} $ or $ v_{n,\sigma}.$ After some simple algebra, the dispersion equation for the mode occurring in the window of transparency specified by the first inequality (\ref{26}), accept the form:
   \be
 1 + \frac{\hbar\Omega}{4t} \frac{w_{n,+}}{w_{n,+}^2 - u^2} + \frac{\hbar\Omega}{4t} \frac{w_{n+1,-}}{w_{n+1,-}^2 - u^2} = 0   \label{28}
    \ee
  Similarly, we obtain the dispersion equation for the mode, which may appear in the window, determined by the second inequality  (\ref{26}):
    \be
  1 + \frac{\hbar\Omega}{4t} \frac{v_{n,-}}{u^2 - v_{n,-}^2} + \frac{\hbar\Omega}{4t} \frac{v_{n+1,+}}{u^2 -v_{n+1,+}^2}  = 0 .  \label{29} 
  \ee
  The solutions of these equations corresponding to the weakly damped longitudinal modes are shown in the Fig. 3.  The curves plotted in this figure represent modes related to the maximum and minimum cross-sections of the Fermi surface, respectively. They are picked out of the sets of branches, each branch corresponding to a certain value of the Landau number.

  \begin{figure}[t]  
\begin{center}
\includegraphics[width=4.6cm,height=9cm,angle=-90]{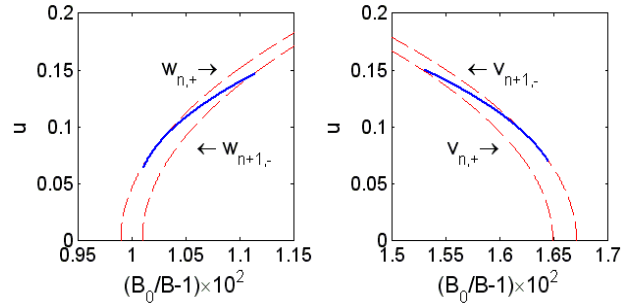}
\end{center}
\caption{
(Color online) Magnetic field dependencies of slow longitudinal quantum modes propagating along the magnetic field related to the maximum (left panel) and minimum (right panel) cross-sections of a Q2D Fermi surface. The speeds of waves shown in the figure are close to the reduced longitudinal velocities of the charge carriers $ w_{n+1,-},\ w_{n,+}, $ (left panel) and $  v_{n,+}, \ v_{n+1,-} $ (right panel). The curves are plotted using the same values of relevant parameters as in the Fig. 1.
}%
 \label{rateI}%
\end{figure}

Low frequency modes, whose speeds have the same order as the speed of sound waves propagating along the axis of the Q2D Fermi surface may be excited by the latter. In the next Section we consider the interaction of quantum waves with the sound and its effect on quantum oscillations of the sound speed.

\subsection{\normalsize III. Interaction of the longitudinal quantum waves with sound waves}

Elastic response of a metal to an external deformation includes rearrangements in the system of conduction electrons. When a sound wave travels in the metal, it gives rise to a collective motion of the charge carriers producing a self-consistent electromagnetic field accompanying the wave. Also, the lattice deformation gives rise to corrections to the crystalline fields, which could be described by means of deformation potentials. However, in the considered case accounting for these corrections does not bring any qualitative changes in the results, so we omit them for brevity.  The electron contribution to the elastic response of the metal manifests itself in magnetic oscillations of sound velocity and  attenuation, and in the coupling of electromagnetic waves, occurring in the metal in the presence of an external magnetic  field, to the sound waves of fitting polarizations and propagation directions. Here, we consider interaction of the longitudinal quantum waves traveling along the symmetry axis of the Fermi surface of a Q2D conductor with the longitudinal sound waves propagating in the same direction. Applying  general equations for the magnetoacoustic response of a metal \cite{13,14}, we may write the following equation for the lattice displacement vector $ {\bf u}_l = (0,0,u_l): $
   \be
    \rho_m \frac{\partial^2 u_l}{\partial t^2} = \lambda \frac{\partial^2u_l}{\partial z^2}  + QE      \label{30}
     \ee 
 where $ \rho_m,\ \lambda $ are the material density and the appropriate elastic constant of the lattice, respectively; $ Q $ is the charge density of the lattice, and $ {\bf E } = (0,0,E) $ is the electric field accompanying the lattice displacements  due to the charge carriers response. The field $ \bf E $ obeys  the Maxwell equations, which gives:
   \be
 \frac{1}{c^2} \frac{\partial^2E}{\partial t^2} = \frac{4\pi}{c} \left(J_z + Q \frac{\partial u_l}{\partial t} \right).   \label{31}
  \ee
  Here, the term $ Q \partial u_l/\partial t $ corresponds to the lattice contribution to the electric current, whereas $ J_z(\sigma,t) $ is the contribution from the charge carriers. 
  Now, we assume
that the time and space dependencies  of the lattice displacement $ u_l $ and the electric field $ E $ have a simple harmonic form, namely: $ u_l = u_{q\omega} \exp (iqz - i\omega t),\ E = E_{q\omega}\exp(iqz - i\omega t). $       As follows from Eqs. (\ref{30}), (\ref{31})
  \be
 E_{q\omega} = \frac{q^2Qu_{q\omega}}{e^2g_0} (Y_+ + Y_-)^{-1}   \label{32} 
 \ee
  where $ g_0 $ is the charge carriers density of states on the Fermi surface in the absence of the  magnetic field and  $ Y_\pm $ are introduced by the Eqs. (\ref{15}). Using Eqs. (\ref{30})-(\ref{32}) we may derive the dispersion equation for the coupled quantum waves and the sound:
   \be
 \omega^2 = q^2s_0^2 + \frac{q^2 N^2}{\rho_m g_0 (Y_+ + Y_-)} .   \label{33}
  \ee
  In this equation, $ s_0 = \sqrt{\lambda/\rho_m} $ is the ``lattice contribution" to the speed of sound, and $ N $ is the charge carriers density. The equation (\ref{33}) agrees with the corresponding result obtained for the particular case of an  isotropic metal with a spherical Fermi surface \cite{8}. From this equation one could see that when the denominator of the second term on the right hand side turns zero indicating the quantum wave appearance, this strongly affects the sound frequency. Thus,  the dispersion equation (\ref{33}) gives a positive evidence of interaction between the sound and quantum waves.

We can analytically solve Eq. (\ref{33}) in the long wavelength limit $ qR\ll 1$ provided that the spin splitting energy $ \ds \frac{g}{2} \hbar \Omega_0 $ equals the cyclotron quantum $ \hbar \Omega. $ In this particular case the longitudinal velocities of the charge carriers associated with two subsequent values of the Landau number and different spin orientations coincide. For certainty, we assume $ w_{n,+} = w_{n+1,-} \equiv w_n, $ and we analyze the solutions of the Eq. (\ref{33})  with $ u $ close the $ w_n. $ We approximate:
  \be
 \eta_\pm \approx \frac{\pi u}{4} \cot \left [\frac{\pi t}{\hbar\Omega} (w_n^2 - u^2) \right] \approx \frac{\hbar\Omega}{4t} \frac{w_n}{w_n^2 - u^2} . 
  \label{34}
        \ee
   Using this approximation,  (\ref{33})  could be reduced to a biquadratic equation, whose solutions have the form:
  \be
 u^2 = \frac{1}{2}(u_0^2 + U_n^2) \pm \frac{1}{2} \sqrt{(u_0^2 - U_n^2)^2 + 4U_n^2 s_1^2}.      \label{35}
 \ee
  Here,$ u_0 = s/v_0,\ s = \sqrt{s_0^2 + s_1^2} $ is the speed of sound including the contribution from the charge carriers $ s_1 = \sqrt{N^2/\rho_m g_0}, $ and $ U_n $ is the speed of the quantum wave determined by the expression:
  \be
 U_n^2 = w_n^2 + \frac{\hbar\Omega}{2 t} w_n.\label{36}
    \ee

The dispersion equation solutions of the form (\ref{35}) are shown in the left panel of the Fig. 4. The plotted curves obviously correspond to  the coupled  sound and electromagnetic quantum waves of the same polarization and propagation direction. In more practical situation when $ \ds \frac{g}{2}\hbar\Omega_0 \neq \hbar\Omega, $ the dispersion equation (\ref{33}) describes coupled quantum and sound waves, as well.

The interaction with the quantum waves makes a significant effect on the magnetic quantum oscillations of the the sound velocity. When the temperature is moderately low $(u^{-1} > \hbar\Omega/kT > 1) $ and the charge carriers scattering is moderately weak $ (1 < ql < t/\hbar\Omega),$ the oscillations of the functions $ \eta_\sigma $ are small in magnitude. As shown in Ref. \cite{15}, in this regime the oscillating correction to the velocity of sound includes two terms, one being proportional to $ \eta_+ + \eta_-$ and another one to $ \Delta_+ + \Delta_-. $ These terms describe magnetic oscillations different in shape, phase and period but their magnitudes are of the same order. However, at very low temperatures and weak scattering, when the inequalities (\ref{23}) are satisfied, the peak values of the functions $ \eta_\sigma $ may reach and exceed unity. At
first glance, this gives grounds to conjecture that in this case the magnetic quantum oscillations of the sound velocity would be significantly increased in magnitude.  However, in the regime specified by Eqs. (\ref{23}), the propagation of the sound wave is accompanied by generation of the sequence of quantum waves branches. This is shown in the right panel of the Fig. 4. Tracing the sound branch in this figure, we see that jumps in the sound velocity related to the interaction with the quantum waves are rather small. Their magnitudes significantly concede the sound velocity value $ s $ taken in the absence of an external magnetic field.
  
   \begin{figure}[t]  
\begin{center}
\includegraphics[width=4.6cm,height=9cm,angle=-90]{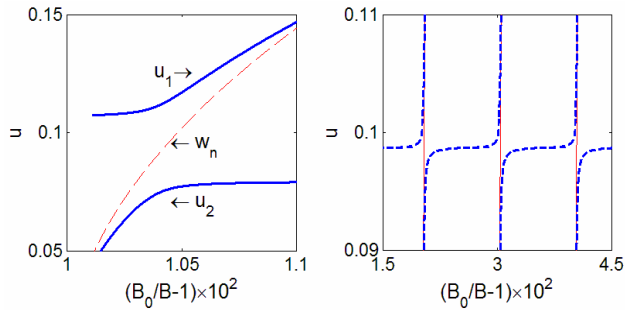}
\end{center}
\caption{
(Color online) Coupling of  slow longitudinal quantum waves to the longitudinal sound propagating along the magnetic field in a Q2D metal. The quantum mode is related to the maximum cross-sectional area of the Fermi surface. The curves are plotted assuming that the relevant parameters accept the same values as in the previous figures. 
}%
 \label{rateI}%
\end{figure}

Finally, we briefly consider magnetic field dependence of the longitudinal component of the charge carriers spin density $ s_z = s_{q\omega} \exp (iqz -i\omega t) $ under the conditions determined by Eqs. (\ref{23}). The spin density amplitude $ s_{q\omega} $is proportional to the difference between the densities of the charge carriers with different spin orientations, so we may write:
  \be
s_{q\omega} = \frac{1}{2} (n_{q\omega}^+ - n_{q\omega}^-).        \label{37}
  \ee
  Here, the Fourier transform of the charge carriers density is described by the expression:
 \be
n_{q\omega}^\sigma = \sum_{\gamma\gamma'} \rho_{\nu\nu'} (\omega,{\bf q}) n_{\nu'\nu} \bf (q)  \label{38}
    \ee
  where $ \gamma $ is the set of the orbital quantum numbers$(\{\nu\} = \{\gamma,\sigma\}) $ and the density matrix $ \rho$ is detrmined by the kinetic equation (\ref{8}). As follows from (\ref{38}), the charge carriers density Fourier transform is proportional to $ E_{q\omega}:$
  \be
 n_{q\omega}^\pm = \frac{ie^2g_0}{q} E_{q\omega} Y_\pm.     \label{39}
  \ee
 Combining this result with the expression (\ref{32}) we get:
   \be
  n_{q\omega}^\pm = in q Q u_{q\omega} \frac{Y_\pm}{Y_+ + Y_-}    \label{40}
   \ee
  So, we obtain:
  \be
 s_{q\omega} = \frac{1}{2} i q Q \frac{Y_+ - Y_-}{Y_+ + Y_-} u_{q\omega}  \label{41}
  \ee
  This shows that  the quantum waves  generated by combined effects of the lattice vibrations and the collective motions of charge carriers providing the electroneutrality of the whole system,  are accompanied by  magnetic oscillations of the longitudinal spin density.

\subsection{\normalsize IV. Concluding remarks}

In the present work we theoretically analyzed longitudinal collective modes which could propagate along the quantizing magnetic field in Q2D conductors. Within the chosen geometry the magnetic field was directed at the right angle to the conducting layers. We showed that these ``quantum waves" may occur at low temperatures and weak charge carriers scattering provided that the Fermi surface warping is rather pronounced $(t > \omega p_0/q). $ Especially interesting is the case of slow quantum waves whose speeds are close to the speed of sound propagating in a Q2D metal perpendicularly to the conducting planes. Such slow modes could interact with the sound waves, and this interaction significantly affects magnetic quantum oscillations of the sound velocity. As was mentioned above, the possible appearance of longitudinal quantum waves in conventional quasi-isotropic metals was theoretically predicted a few decades ago but they were not observed in experiments. Q2D conductors provide much better opportunities for such observations. Now, we briefly discuss which properties of Q2D conductors give them the advantage over conventional metals in this respect. The conditions on the temperature and the intensity of charge carriers scattering in quasi-isotropic metals were presented in Ref. \cite{14}. They could be written in the form (\ref{23}) provided that transfer integral $ t $ in the second inequality is replaced by the charge carriers chemical potential $ \mu. $ The restrictments put upon the scattering intensity does not significantly diverse for 3D and Q2D materials for the parameter $ p_0 $ usually takes on values of the same order as the Fermi 
 momentum in a quasi-isotropic metal. However, a pronounced difference reveals itself in the  temperature regime required to provide the quantum waves occurrence in Q2D and 3D metals. Due to the specifics of the charge carriers dispersions in Q2D materials the maximum charge carriers velocity in the direction perpendicular to the conducting layers is much smaller than typical Fermi velocities of conduction electrons in  quasi-isotropic metals. Assuming that quantum waves are traveling with speeds close to the speed of sound propagating along the same direction, the parameter $ u $  in Q2D conductors probably takes on values no less than $ 0.1, $ whereas in 3D metals it has the order of $ 10^{-3} - 10^{-4}.$ Therefore, within the chosen geometry the considered quantum waves may occur in Q2D conductors at temperatures significantly exceeding those required for them to appear in 3D metals. As shown in this work, one may expect to observe the longitudinal quantum waves traveling along the magnetic field in Q2D organic metals and their effect on the magnetic quantum oscillations of the sound velocity at quite accessible temperatures of the order of $ 10 mK. $ Such observations should provide better insight in the specific features of low temperature magnetic quantum oscillations of the elastic response of Q2D conductors.
  \vspace{2mm}

{\bf Acknowledgments:} 
Author thank E. Mele for helpful discussion and G. M. Zimbovsky for help with the manuscript. This work was partly supported  by  NSF-DMR-PREM 0353730.

\subsection{\normalsize \bf V. Appendix}

Expressions for the functions $Y_\pm, $ which appear in the Eqs. (\ref{15}) and (\ref{19})
  are given by:
   \begin{align}
  Y_\sigma = &\frac{1}{4\pi^2\hbar\lambda^2} \frac{1}{g_0}  \sum_n \int  d p_z
  \nn \\  & \times 
 \frac{f(n,p_z,\sigma) - f(n,p_z - \hbar q,\sigma)}{E(n,p_z-\hbar q,\sigma) - E(n,p_z,\sigma) + \hbar\omega + i\hbar/\tau}       \label{42}
   \end{align}                
  where $\lambda $ is the magnetic length. 

Using the Poisson summation formula 
  \be
 \sum_{n=0}^\infty \varphi \left (n + \frac{1}{2} \right) = 
\sum_{r = - \infty}^\infty \varphi (x)  \exp(2\pi irx)          \label{43}
  \ee
 we may present the functions $ Y_\sigma $ as sums of terms 
$ \ds  Y_{0\sigma} = \frac{m_\perp p_0}{2\pi^2 \hbar^3} \frac{1}{g_0} \Big(1 + \frac{i\pi u}{2} \Big) \equiv \frac{a}{2} \Big(1 + \frac{i\pi u}{2}\Big) $    independent of the magnetic field magnitude and oscillating contributions $ \tilde Y_\sigma. $ The latter could be approximated by  expressions:
   \be
 \tilde Y_\sigma = \tilde Y_\sigma (q) + \tilde Y_\sigma(-q)          \label{44}
 \ee
 where
  \begin{align}
 \tilde Y_\sigma (\pm q) = & \frac{m_\perp}{(2\pi\hbar)^2} \frac{1}{g_0} \sum_{r=1}^\infty (-1)^r \int_0^\infty f(E, \sigma) dE
   \nn \\  \times &
 \int_{-p_0}^{p_0} \frac{\cos [2\pi r F(E,p_z,\sigma)/B] d p_z}{E(n,p_z \pm \hbar q,\sigma) - E(n,p_z,\sigma) \mp \hbar\omega \mp i\hbar/\tau} .           \label{45}
 \end{align}
   Here, 
  \be
  F(E,p_z,\sigma) = \frac{c}{2\pi\hbar e} A(E,p_z,\sigma)   \label{46}
   \ee
 and $A(E,p_z,\sigma) $ is the cross-sectional area of the corresponding isoenergetic surface.

 \begin{figure}[t]  
\begin{center}
\includegraphics[width=4.8cm,height=8cm,angle=-90]{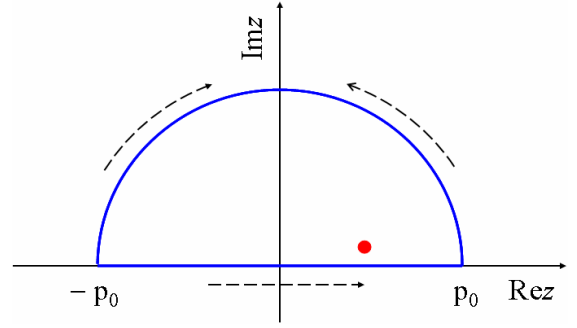}
\end{center}
\caption{
(Color online)  The schematic of the contour chosen to carry out integration over $ p_z $ in the expression (\ref{45}). The point indicates a pole in the integrand. 
}%
 \label{rateI}%
\end{figure} 
  
To properly estimate the integrals over $ p_z $ in Eq. (\ref{45}) we extend integrands over the upper halves of the complex plane 
 and we choose the integration paths consisting of  segments of the real axes $ - p_0 \leq p_z \leq p_0 $  and  semicircles of the radii $ p_0 $  (see Fig. 5). The integrands possess poles, whose positions are determined  by equations:
  \be
 \sin\left(\pi \frac{p_z}{p_0}\right) \pm \frac{\pi\hbar q}{2p_0} \cos\left(\pi \frac{p_z}{p_0}\right) - u - \frac{i}{ql} = 0   \label{47}
  \ee
  where $ u = \omega/qv_0,\ l = v_0\tau, $ and $ v_0 = 2\pi t/p_0 $ is the maximum value of the charge carriers velocity component along the magnetic field. The poles are situated 
within the contours of integration when the Fermi surface corrugation is sufficiently pronounced to satisfy the inequality $ t > \omega p_0/q. $ 
 The residues from these poles give contributions to the oscillating parts of the functions $ Y_\sigma $. Besides, the oscillating parts include corrections originating from the integration over the semicircular arcs. Assuming that $ u \ll 1 $ we obtain the following approximations for $ Y_\sigma: $ 
  \be
 Y_\sigma = \frac{a}{2} \Big(1 + \eta_\sigma + \Delta_\sigma + \frac{i\pi u}{2} + i \Gamma_\sigma \Big)  \label{48}
 \ee
Here, $ \eta_\sigma $ and $ \Gamma _sigma $ are equal to the real and imaginary  parts of the function $ X_\sigma, $ respectively. The function $ X_\sigma $ has the form (\ref{17})
Neglecting for a while the effects of temperature and scattering, we may carry out summation over $``r" $ in the expression (\ref{17}),      using the formula \cite{19}:
  \be
  \sum_{k=1}^\infty (-1)^{k-1} \frac{\cos(kx)}{k} = \ln \Big\{2\cos\frac{x}{2} \Big\} .   \label{49}
  \ee
We get
  \begin{align}
\eta_\sigma = & \frac{1}{4qR} \ln 
\frac{\ds \cos \Big[\pi \Big(\frac{F_{\max}}{B} -\frac{\sigma g\Omega_0}{\Omega} + \frac{2t}{\hbar\Omega} \Big(\sqrt{1 - U_-^2} - 1\Big) \Big) \Big]}{\ds \cos \Big[\pi \Big(\frac{F_{\max}}{B} -\frac{\sigma g\Omega_0}{\Omega} + \frac{2t}{\hbar\Omega} \Big(\sqrt{1 - U_+^2} - 1\Big) \Big) \Big]} 
  \nn \\  + &
 \frac{1}{4qR} \ln 
\frac{\ds \cos \Big[\pi \Big(\frac{F_{\min}}{B} + \frac{\sigma g\Omega_0}{\Omega} -\frac{2t}{\hbar\Omega} \Big(\sqrt{1 - U_-^2} - 1\Big) \Big) \Big]}{\ds \cos \Big[\pi \Big(\frac{F_{\min}}{B} + \frac{\sigma g\Omega_0}{\Omega} - \frac{2t}{\hbar\Omega} \Big(\sqrt{1 - U_+^2} - 1\Big) \Big) \Big]}  
           \label{50}
  \end{align}
  where $ F_{\min}  $ and $ F_{\max} $ correspond to the minimum and maximum cross-sectional areas of the Fermi surface.

When $ U_\pm $ approaches one of the dimensionless velocities given by the Eqs. (\ref{20}), (\ref{21}), the value of the corresponding cosine  in the Eq. (\ref{50}) approaches zero, and the function $ \eta_\sigma $ diverges. For weakly corrugated FS $(t < \omega p_0/q), $ the pole is situated without the contour of integration, therefore the terms $ \eta_\sigma $ and $ \Gamma_sigma $ disappear from the expression for $ Y_sigma. $ Besides the contributions from the pole, the expression (\ref{48}) includes the terms $ \Delta_\sigma $ originating from the integration over the semicircular are shown in the Fig. 5.  At $ t/\hbar \omega > 1 $ these terms may be presented as follows \cite{14}:
  \begin{align}
  \Delta_\sigma = & \frac{1}{4} \sqrt{\frac{\hbar\Omega}{2\pi^2 t}} \sum_r \frac{(-1)^2}{\sqrt r} D(r)
 \Big\{ \exp \Big[\frac{2\pi ir F_{max}^\sigma}{B} - \frac{i\pi}{4} \Big] G_{r\sigma}^- 
 \nn\\ & +
 \exp \Big[- \frac{2\pi ir F_{min}^\sigma}{B} + \frac{i\pi}{4} \Big] G_{r\sigma}^+ \Big\}     \label{51}
 \end{align} 
  where 
 \be
 G_{r\sigma}^\pm = \pm \frac{i}{2} \int_0^\infty \exp[\pm iy] \exp\Big[-2\pi \sqrt{{rF_{max/min}^\sigma u^2y}\big/{B}}\Big] dy   . \label{52}
  \ee
  The ratio $ F/B $ has the same order as $ \mu/\hbar\Omega $ where $ \mu $ is the charge carriers chemical potential and $ \hbar \Omega $ is the cyclotron quantum. Under the condition $ u \sqrt{\mu/\hbar\Omega} > 1 $ we may approximate the functions $ G^\pm $ using asymptotic expressions for Fresnel integrals, and we get $ G_r^\pm \approx 1. $ Substituting these approximations into the Eq. (\ref{51}) we obtain:
       \begin{align}
   \Delta_\sigma = & \sqrt{\frac{\hbar\Omega}{2\pi^2t}} \sum_{r=1}^\infty 
\frac{(-1)^r}{\sqrt r} D(r) 
      \nn \\  & \times
  \left\{\cos\left[\frac{2\pi r F_{max}}{B} - \pi r \sigma \frac{ g}{2} \frac{\Omega_0}{\Omega} - \frac{\pi}{4} \right] \right.
  \nn \\ & +  \left.
\cos\left[\frac{2\pi r F_{min}}{B} + \pi r \sigma \frac{g}{2} \frac{\Omega_0}{\Omega} + \frac{\pi}{4} \right] \right\}.   \label{53}
   \end{align}


\end{document}